\documentclass[11pt]{article}

\usepackage[utf8]{inputenc}
\usepackage[T1]{fontenc}
\usepackage{amsmath}
\usepackage{amssymb}
\usepackage{authblk}
\usepackage[numbers,sort&compress]{natbib}
\usepackage[colorlinks=true, urlcolor=blue, citecolor=blue, linkcolor=blue]{hyperref}
\usepackage{enumitem}
\usepackage{xcolor}
\usepackage{graphicx}

\title{\texttt{SALTED}: a symmetry-adapted machine-learning program for predicting electron-densities in molecules and materials}

\author[1]{Zekun Lou\thanks{ORCID: 0009-0009-7792-3202}}
\author[2]{Alan M. Lewis\thanks{ORCID: 0000-0002-3296-7203}}
\author[5]{Théophane Bernhard\thanks{ORCID: 0009-0001-0130-5012}}
\author[3]{Lukas Seifert\thanks{ORCID: 0009-0009-4032-058X}}
\author[6]{Agustin Salcedo\thanks{ORCID: 0000-0001-5525-8605}}
\author[3]{Florian Kleemiss\thanks{ORCID: 0000-0002-3631-1535}}
\author[1,4]{Mariana Rossi\thanks{ORCID: 0000-0002-3552-0677}}
\author[5]{Andrea Grisafi\thanks{ORCID: 0000-0003-1433-125X. Corresponding author.}}

\affil[1]{MPI for the Structure and Dynamics of Matter, Hamburg, Germany.}
\affil[2]{Department of Chemistry, University of York, York, UK.}
\affil[3]{Institute of Inorganic Chemistry, RWTH Aachen University, Landoltweg 1a, 52074 Aachen, Germany.}
\affil[4]{Yusuf Hamied Department of Chemistry, Cambridge University, Cambridge, UK.}
\affil[5]{Physicochimie des Électrolytes et Nanosystèmes Interfaciaux, Sorbonne Université, CNRS, F-75005 Paris, France.}
\affil[6]{Laboratoire de physique de L’École normale supérieure de Paris, CNRS, ENS \& Université PSL, Sorbonne Université, Université de Paris, F-75005 Paris, France.}

\date{}

\begin{document}

\maketitle
\newpage
\section*{Summary}

\texttt{SALTED} provides an open-source Python package for machine learning the quantum-mechanical electron density, $n(\mathbf{r})$, in molecular and condensed-phase systems based on input atomic coordinates and species~\cite{gris+19acscs,lewis+21jctc,Grisafi2023}. The program adopts a linear atom-centered decomposition of the electron density, which makes it highly transferable across diverse atomistic configurations sharing similar chemical environments. Because of this representation choice, \texttt{SALTED} is naturally interfaced with state-of-the-art electronic-structure programs based on atomic orbitals, namely CP2K~\cite{cp2k-made-easy-2026}, FHI-aims~\cite{roadmap-aims-2026}, and PySCF \cite{PySCF2020}, from which reference electron-density data can be generated and used to train a model. The learning algorithm is based on a symmetry-adapted extension of Gaussian process regression \cite{gris+18prl}, making \texttt{SALTED} especially efficient in small-data regimes. Thanks to the implementation of vector-field kernel functions~\cite{Rossi2025}, \texttt{SALTED} can also learn the first-order response of the electron density to applied electric fields, $\partial n(\mathbf{r})/\partial \mathbf{E}$. The application of \texttt{SALTED} within computational workflows has already shown its utility in a wide variety of contexts, including the calculation of polarization vectors~\cite{grisafi2023prm} and polarizability tensors~\cite{Rossi2025}, the accurate evaluation of Coulomb forces in QM/MM molecular-dynamics simulations~\cite{Grisafi2024}, and electronic-structure studies of large-scale 2D materials~\cite{lou2026prx}. 

 \section*{Statement of need}

Electronic structure methods can be predictive for the simulation of the properties of a wide variety of materials. Among these methods, density-functional theory (DFT) has, perhaps, become the most successful and popular, having solved outstanding problems in areas as diverse as biochemistry, catalysis, nanotechnology and quantum materials~\cite{Burke2012persp,Jones2015rmp}. The popularity of DFT is based on its ability to provide sufficient accuracy for many practical applications at a modest cost. However, the cost of these calculations can only be considered modest in comparison to other, more advanced, electronic structure methods. For the current landscape of data-driven material science and molecular discovery, allied to the quest of achieving first-principles accuracy for larger systems and longer time scales, even the cost of DFT calculations becomes prohibitively large.

A machine-learning method that can predict the most fundamental quantity of DFT, namely the real-space electronic density, therefore holds immense potential to deliver a single model that can be used to calculate a multitude of downstream material properties at a small fraction of the cost of DFT, while maintaining very similar accuracy. In this paper, we describe the \texttt{SALTED} software package, which allows electronic density and electronic-density response predictions over an atomic basis consistent with the underlying electronic-structure architectures.

The availability of this open-source software package provides the atomistic-simulation community with a practical and extensible framework for constructing machine-learning models of the electron density and its linear response from first-principles reference data. \texttt{SALTED} is intended for researchers in computational chemistry, condensed-matter physics, materials science, and molecular simulation who wish to accelerate electronic-structure calculations while retaining direct access to physically meaningful electronic observables. By interfacing with widely used electronic-structure packages, \texttt{SALTED} can be embedded into established computational workflows. The package is designed both for users seeking efficient predictions of electron densities and derived properties, and for method developers interested in extending the methodology or in interfacing density predictions with downstream electronic-structure and multiscale simulation tools. \texttt{SALTED} facilitates reproducible research and supports the development of machine-learning models for electronic structure by providing an open, documented, and reusable implementation of electron-density prediction. 

\section*{State of the field}

Two main directions have been independently followed to construct ML models of the electron density that either directly sample $n(\mathbf{r})$ on a real-space 3D grid, or expand $n(\mathbf{r})$ over a linear atomic basis. Examples of grid-based ML models with published open-source packages are: \texttt{DeepDFT}, based on equivariant graph neural networks~\cite{Jorgensen2022}; a linear-regression framework based on Jacobi–Legendre many-body descriptors~\cite{Focassio2023}; \texttt{Charge3net}, based on a high-order equivariant neural network~\cite{Koker2024}; and an extension of \texttt{FIREANN}, where an efficient sampling strategy is devised to drastically reduce the number of grid-point evaluations and to further treat the response to applied fields~\cite{Feng2025}. Albeit not targeting directly the electronic density, the Materials Learning Algorithm (\texttt{MALA}) package~\cite{Cangi_MALA_2025} learns the local density of states on a grid, and reconstructs $n(\mathbf{r})$ from this quantity. 

When compared with grid-based approaches, ML models that represent the density on an atomic basis carry the advantage of greatly reducing the amount of data at the price of an acceptably small error due to the finite expressiveness of the basis functions. The \texttt{SALTED} package is the result of one of the earliest ML approaches developed to learn the electron density via a set of atomic coefficients~\cite{gris+19acscs}. Specifically, $n(\mathbf{r})$ is expanded over a linear basis made of radial functions $R^{\lambda}_{n}$ and spherical harmonics $Y_{\mu}^{\lambda}$ centered around each atom of the system:
\begin{equation}\label{eq:rhoexp}
    n(\mathbf{r}) = \sum_{in\lambda\mu} c_{in\lambda\mu} \sum_{\mathbf{u}} R^\lambda_{n}(\left|\mathbf{r}-\mathbf{r_{i}} -\mathbf{u}\right|)\, Y_{\mu}^{\lambda}\left(\widehat{\mathbf{r}-\mathbf{r_{i}}-\mathbf{u}}\right)
\end{equation}
where $i$ are the atomic indexes, $\boldsymbol{u}$ are the cell translation vector when considering periodic systems~\cite{lewis+21jctc}, and $c_{in\lambda\mu}$ are the density expansion coefficients. The reference data for $c_{in\lambda\mu}$ can directly be obtained from resolution of the identity (RI) methods implemented in electronic-structure codes that are based on atomic orbitals, making \texttt{SALTED} easy to  interface with codes using such basis sets and methods. 

As a key difference from the models previously mentioned, \texttt{SALTED} is based on a symmetry-adapted extension of Gaussian process regression (GPR). Spherical-tensor kernels~\cite{gris+18prl}, $\mathbf{K}^\lambda$, rather than equivariant neural-networks, are computed to satisfy exact $\mathcal{O}(3)$ symmetries~\cite{gris+19acscs}, starting from atom-density features obtained using the \texttt{featomic}  library~\cite{bigi_metatensor_2026}. While the adoption of a GPR method limits the scalability of \texttt{SALTED} over strongly heterogeneous datasets~\cite{Grisafi2023}, the constrained function space defined by the kernels, together with the physically interpretable nature of the underlying atomistic features, helps avoid overfitting---thereby facilitating transferable extrapolations of the electron density across different system sizes. Moreover, when compared with data-hungry neural-network architectures, the data-efficiency of GPR-based approaches becomes a great advantage when training the model on high-level electronic-structure methods, for which only a small number of calculations ($\sim 10^{2-3}$) can be performed.

In addition, \texttt{SALTED} presents a couple of distinctive features. First, a suitable linear combination of the kernels $\mathbf{K}^\lambda$ allows \texttt{SALTED} to learn the first-order response of the density to uniform applied electric fields, $\partial n(\mathbf{r})/\partial E_k$, thereby ensuring the correct transformations of a 3D vector field~\citep{Rossi2025}.
Second, the use of the \texttt{featomic} library enables the inclusion of long-distance equivariants (LODE) features~\citep{grisafi2019jcp}, which are essential for learning the nonlocal redistribution of $n(\mathbf{r})$ in highly-polarizable systems such as metallic frameworks \citep{grisafi2023prm,Rossi2025}.

We note that there are other ML packages that share the same philosophy as \texttt{SALTED} in representing the electron density similarly to Eq.~\eqref{eq:rhoexp}, albeit with different features and capabilities: equivariant neural-network architectures based on the \texttt{E3NN} model~\cite{Rackers2023};   \texttt{scdp}, which augments the atom-centered basis with additional ``virtual'' orbitals placed at non-atomic sites (e.g., bond midpoints) to improve expressivity, and regresses the expansion coefficients with a high-capacity equivariant neural network~\cite{fu2024recipe}; and lastly, an equivariant neural-network architecture that predicts atom-centered density coefficients as an intermediate representation for learning energies and forces~\cite{bogojeski2026}.

\section*{Software design}

\begin{figure}[t!]
    \centering
    \includegraphics[width=1\linewidth]{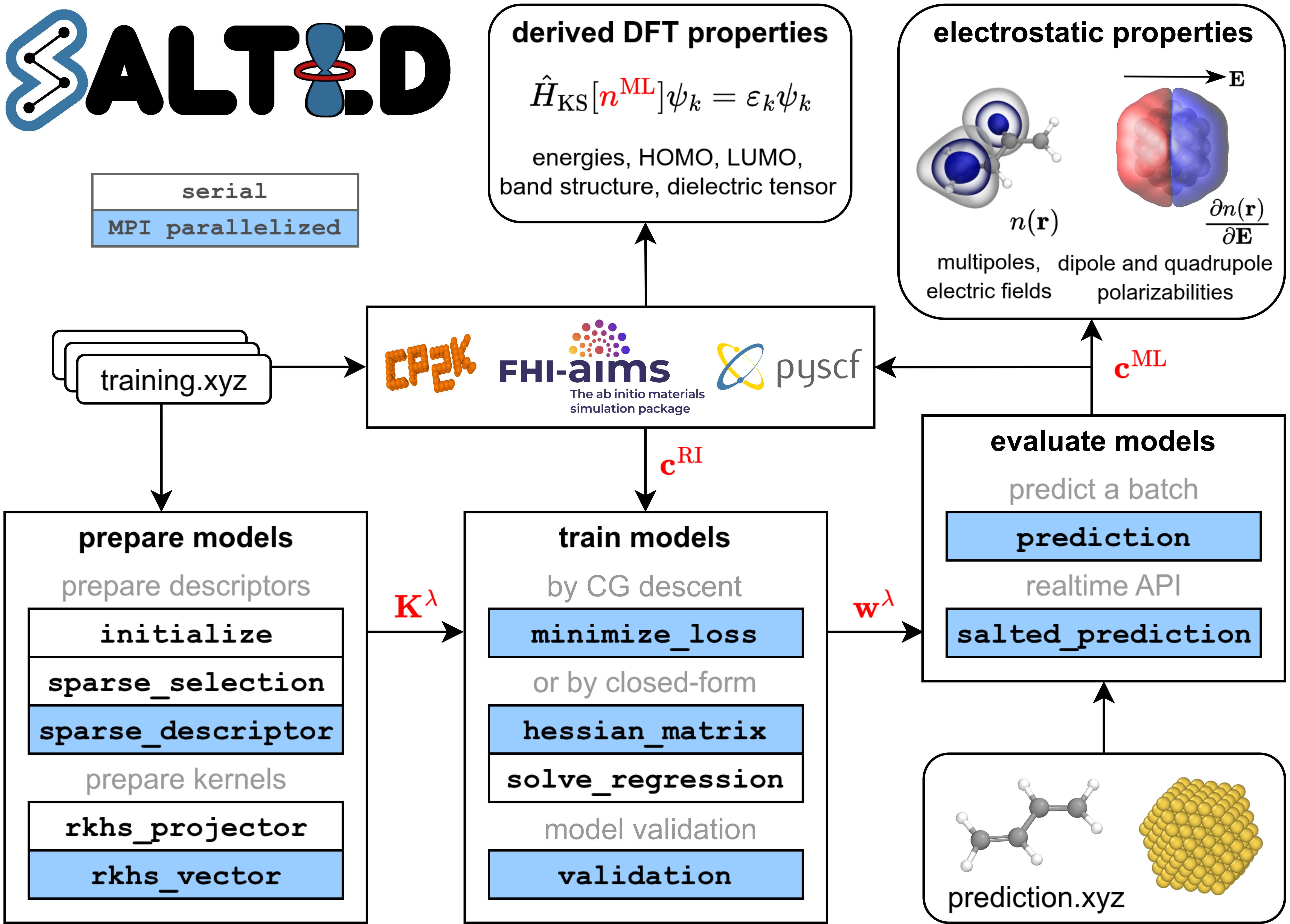}
    \caption{
        \textbf{\texttt{SALTED} workflow}.
        Reference density or density-response RI coefficients \(\mathbf{c}^\mathrm{RI}\) of the training dataset are generated by \texttt{CP2K}, \texttt{FHI-aims}, or \texttt{PySCF}.
        Model preparation builds \(\lambda\)-descriptors and assembles kernels \(\mathbf{K}^{\lambda}\) of the training structures, with optional sparsification.
        Model training obtains regression weights \(\mathbf{w}^\lambda\) using \(\mathbf{c}^\mathrm{RI}\) and \(\mathbf{K}^\lambda\).
        The trained models can be quickly validated to determine their accuracy, before being used to evaluate unseen structures.
        Predicted RI coefficients \(\mathbf{c}^\mathrm{ML}\) are used either directly, to evaluate electrostatic properties such as multipoles, electric fields, and polarizabilities, or as an input density from which the Kohn--Sham Hamiltonian is built and further obtain DFT properties after one diagonalization step.
        Shaded boxes mark stages that are parallelized by MPI.
    }
    \label{fig:workflow}
\end{figure}

\texttt{SALTED} is organized as a workflow in three stages: dataset preparation, model training, and prediction (Figure~\ref{fig:workflow}).
Each stage is configured through a dedicated section of a single \texttt{inp.yaml} file, and each step within a stage is exposed as an independently invocable command, \texttt{python -m \texttt{salted}.<step>}.
GitHub Continuous Integration (CI) is enabled for testing the package: unit tests cover the utility modules, and pipeline tests run the full workflow through every electronic-structure interface.

This step-wise organization allows a run to be check-pointed, restarted, and distributed across separate HPC allocations, with the intermediate quantities of each step available on disk.
Such control is necessary in practice: training a capable model over a large chemical space or on condensed-phase systems usually requires many HPC nodes, and the most efficient parallelization strategy differs from step to step.
Users can resume an interrupted run, reuse descriptors across hyperparameter choices, and inspect intermediate quantities to improve models.
We accept a less immediate interactive experience that a single-call library API would provide in exchange for workflows that remain manageable by the user.

\texttt{SALTED} integrates with a variety of electronic-structure codes based on atomic orbitals, \texttt{CP2K}~\cite{cp2k-made-easy-2026}, \texttt{FHI-aims}~\cite{roadmap-aims-2026}, and \texttt{PySCF}~\cite{PySCF2020}, thus supporting a large fraction of community-developed electronic-structure software. Density data from \texttt{CP2K} are provided on uniform grids as \texttt{*.cube} files following pseudo-valence representations of $n(\mathbf{r})$; \texttt{SALTED} can then compute the required RI coefficients via a dedicated \texttt{density\_fitting} module. 
\texttt{FHI-aims} is an all-electron electronic-structure program; in this case, density data for isolated and periodic systems are directly obtained as RI coefficients on the same footing. Finally, pseudo-valence \texttt{PySCF} density data for finite molecular systems can be obtained from \texttt{SALTED} by calling relevant \texttt{PySCF} functions.

\texttt{SALTED} unifies the training data obtained from these different codes within a single shared format.
This requires extra engineering in the interfaces, but it lets research groups keep the electronic-structure code they already use in production, together with its established settings, and integrate \texttt{SALTED} into an existing workflow rather than switching electronic-structure codes. CI tests cover all the interfaces. We note that both \texttt{CP2K} and \texttt{PySCF} work with Gaussian-type orbitals, enabling straightforward analytical calculations of density and density-response derived properties from the predicted \texttt{SALTED} coefficients, such as Hartree energies, electric-fields, dipoles and polarizabilities. While some of these properties are already computed and printed by \texttt{SALTED} upon density prediction, others could readily be implemented as dedicated post-processing modules.

The performance-critical steps, including equivariant descriptor contraction and the construction of Hessian matrix, are implemented as \texttt{numba}-compiled Python kernels~\cite{numba2015}.
The \texttt{numba} kernels are carefully composed and benchmarked, and their portability does not come at the cost of performance.
MPI parallelization is optional: parallelizable steps run serially by default and use \texttt{mpi4py}~\cite{mpi4py2021} automatically when launched under \texttt{mpirun}.
OpenMP and Python thread pools would parallelize these steps within a single shared-memory node, whereas MPI covers the full range of hardware of the users, from personal computers and small workstations to multi-node HPC clusters. To improve portability, the supplied Dockerfile can be used to build a self-contained image with an MPI-enabled version of \texttt{SALTED}.
\section*{Research impact statement}

The integration of \texttt{SALTED} predictions within computational workflows has already been shown useful in real-case scientific applications. For example, \texttt{SALTED} predictions of the charge-density response in metallic electrodes could be used to achieve a $\sim10^3$ speedup in the calculation of DFT-level electrostatic forces entering the quantum-mechanics/molecular-mechanics simulation of model ionic capacitors \cite{Grisafi2024}. More recently, the use of the predicted density as a DFT input to perform a single Kohn-Sham Hamiltonian diagonalization has enabled the study of the electronic properties of twisted bilayer Moir\'e materials over extremely large supercells \cite{lou2026prx}. Beyond electrostatic and DFT properties, \texttt{SALTED} enables the calculation of electron-density-based features at scale. A relevant example is the application to quantum-chemistry molecular interaction analysis~\cite{fabr+20chimia}. Additionally, the RI density coefficients of Eq.~\eqref{eq:rhoexp} are directly applicable to crystal structure refinement from X-ray and electron diffraction data~\cite{seifert2026}, making the integration of \texttt{SALTED} predictions within these algorithms highly promising.

\section*{AI usage disclosure}

We declare that no AI tool was used to make core code design decisions, nor were they used to make scientific decisions.
Generative AI tools were employed to help with code refactoring and cleaning and with text polishing.

\section*{Acknowledgements}

We acknowledge funding from the French National Research Agency under the France 2030 program (PEPR BATMAN, Grant No. ANR-22-PEBA-0002 and PEPR DIADEM, Grant No. ANR-22-PEXD-0001) and the German Research Foundation (DFG) Project-ID 555467911 - CRC 1772 / TP A06.

\bibliographystyle{unsrt}

\begin{thebibliography}{10}

\bibitem{gris+19acscs}
Andrea Grisafi, Alberto Fabrizio, Benjamin Meyer, David~M. Wilkins, Clemence Corminboeuf, and Michele Ceriotti.
\newblock Transferable machine-learning model of the electron density.
\newblock {\em ACS Cent. Sci.}, 5(1):57--64, 2019.
\newblock URL: \url{https://doi.org/10.1021/acscentsci.8b00551}.

\bibitem{lewis+21jctc}
Alan~M. Lewis, Andrea Grisafi, Michele Ceriotti, and Mariana Rossi.
\newblock Learning electron densities in the condensed phase.
\newblock {\em J. Chem. Theory Comput.}, 17(11):7203--7214, 2021.
\newblock URL: \url{https://doi.org/10.1021/acs.jctc.1c00576}.

\bibitem{Grisafi2023}
Andrea Grisafi, Alan~M. Lewis, Mariana Rossi, and Michele Ceriotti.
\newblock Electronic-structure properties from atom-centered predictions of the electron density.
\newblock {\em J. Chem. Theory Comput.}, 19(14):4451--4460, 2023.
\newblock URL: \url{https://doi.org/10.1021/acs.jctc.2c00850}.

\bibitem{cp2k-made-easy-2026}
Marcella Iannuzzi, Jan Wilhelm, Frederick Stein, Augustin Bussy, Hossam Elgabarty, Dorothea Golze, Anna-Sophia Hehn, Maximilian Graml, Stepan Marek, Beliz~Sertcan G{\"o}kmen, Christoph Schran, Harald Forbert, Rustam~Z. Khaliullin, Anton Kozhevnikov, Mathieu Taillefumier, Rocco Meli, Vladimir~V. Rybkin, Martin Brehm, Robert Schade, Ole Sch{\"u}tt, Johann~V. Pototschnig, Hossein Mirhosseini, Andreas Kn{\"u}pfer, Dominik Marx, Matthias Krack, J{\"u}rg Hutter, and Thomas~D. K{\"u}hne.
\newblock The cp2k program package made simple.
\newblock {\em J. Phys. Chem. B}, 130(4):1237--1310, 2026.
\newblock URL: \url{https://doi.org/10.1021/acs.jpcb.5c05851}.

\bibitem{roadmap-aims-2026}
Joseph~W. Abbott, Carlos Mera~Acosta, Alaa Akkoush, Alberto Ambrosetti, Viktor Atalla, Alexej Bagrets, Joerg Behler, Daniel Berger, Hannah Bertschi, Bj{\"o}rn Bieniek, Jonas Bj{\"o}rk, Volker Blum, Saeed Bohloul, Connor~L. Box, Nicholas~James Boyer, Danilo~Simoes Brambila, Gabriel~A. Bramley, Kyle~R. Bryenton, Mar{\'i}a Camarasa-G{\'o}mez, Christian Carbogno, Fabio Caruso, Sucismita Chutia, Michele Ceriotti, G{\'a}bor Cs{\'a}nyi, William Dawson, Francisco~A. Delesma, Fabio Della~Sala, Bernard Delley, Robert DiStasio, Maria Dragoumi, Sander Driessen, Marc Dvorak, Simon Erker, Ferdinand Evers, Eduardo Fabiano, Matthew~R. Farrow, Florian Fiebig, Jakob Filser, Lucas Foppa, Lukas Gallandi, Alberto Garcia, Ralf Gehrke, Simiam Ghan, Luca Ghiringhelli, Mark Glass, Stefan Goedecker, Dorothea Golze, Matthias Gramzow, James~A. Green, Andrea Grisafi, Andreas Gr{\"u}neis, Johannes~Jan G{\"u}nzl, Stefan Gutzeit, Samuel~J. Hall, Felix Hanke, Ville Havu, Xingtao He, Joscha Hekele, Olle Hellman, Uthpala Herath, Jan Hermann,
  Daniel Hernang{\'o}mez-P{\'e}rez, Oliver~T. Hofmann, Johannes Hoja, Simon Hollweger, Lukas H{\"o}rmann, Benjamin Hourahine, Wei~Bin How, William~P. Huhn, Marcel H{\"u}lsberg, Timo Jacob, Sara Panahian~Jand, Hongbing Jiang, Erin Johnson, Werner J{\"u}rgens, Juhan~Matthias Kahk, Yosuke Kanai, Kisung Kang, Petr Karpov, Elisabeth Keller, Roman Kempt, Danish Khan, Matthias Kick, Benedikt~P. Klein, Jan Kloppenburg, Alexander Knoll, Florian Knoop, Franz Knuth, Simone~S. K{\"o}cher, Jannis Kockl{\"a}uner, Sebastian Kokott, Thomas K{\"o}rzd{\"o}rfer, Hagen-Henrik Kowalski, Peter Kratzer, Pavel Kus, Raul Laasner, Bruno Lang, Bj{\"o}rn Lange, Marcel~F. Langer, Ask~Hjorth Larsen, Hermann Lederer, Susi Lehtola, Maja-Olivia Lenz-Himmer, Moritz Leucke, Sergey Levchenko, Alan Lewis, O.~Anatole von Lilienfeld, Konstantin Lion, Werner Lipsunen, Johannes Lischner, Yair Litman, Chi Liu, Qing-Long Liu, Songrui Liu, Andrew~J. Logsdail, Michael Lorke, Zekun Lou, Iuliia Mandzhieva, Andreas Marek, Johannes~T. Margraf, Reinhard~J.
  Maurer, Tobias Melson, Florian Merz, J{\"o}rg Meyer, Georg~S. Michelitsch, Teruyasu Mizoguchi, Evgeny Moerman, Dylan Morgan, Jack Morgenstein, Jonathan Moussa, Akhil~S. Nair, Lydia Nemec, Harald Oberhofer, Alberto Otero-de-la Roza, Ram{\'o}n~L. Panad{\'e}s-Barrueta, Thanush Patlolla, Mariia Pogodaeva, Alexander P{\"o}ppl, Alastair J.~A. Price, Thomas A.~R. Purcell, Jingkai Quan, Nathaniel Raimbault, Markus Rampp, Karsten Rasim, Ronald Redmer, Xinguo Ren, Karsten Reuter, Norina~A. Richter, Stefan Ringe, Patrick Rinke, Simon~P. Rittmeyer, Herzain~I. Rivera-Arrieta, Matti Ropo, Mariana Rossi, Victor Ruiz, Nikita Rybin, Andrea Sanfilippo, Matthias Scheffler, Christoph Scheurer, Christoph Schober, Franziska Schubert, Tonghao Shen, Christopher Shepard, Honghui Shang, Kiyou Shibata, Andrei Sobolev, Ruyi Song, Aloysius Soon, Daniel~T. Speckhard, Pavel~V. Stishenko, Elia Stocco, Muhammad~N. Tahir, Izumi Takahara, Jun Tang, Zechen Tang, Thomas Theis, Franziska Theiss, Alexandre Tkatchenko, Milica Todorovi{\'c},
  George Trenins, Oliver~T. Unke, {\'A}lvaro V{\'a}zquez-Mayagoitia, Oscar van Vuren, Daniel Waldschmidt, Han Wang, Yanyong Wang, J{\"u}rgen Wieferink, Jan Wilhelm, Scott Woodley, Jianhang Xu, Yong Xu, Yi~Yao, Yingyu Yao, Mina Yoon, Victor Wen-zhe Yu, Zhenkun Yuan, Marios Zacharias, Igor~Ying Zhang, Min-Ye Zhang, Wentao Zhang, Xingchen Zhang, Rundong Zhao, Shuo Zhao, Ruiyi Zhou, Yuanyuan Zhou, and Tong Zhu.
\newblock Roadmap on advancements of the fhi-aims software package.
\newblock {\em Electron. Struct.}, 2026.
\newblock URL: \url{https://doi.org/10.1088/2516-1075/ae8067}.

\bibitem{PySCF2020}
Qiming Sun, Xing Zhang, Samragni Banerjee, Peng Bao, Marc Barbry, Nick~S. Blunt, Nikolay~A. Bogdanov, George~H. Booth, Jia Chen, Zhi-Hao Cui, Janus~J. Eriksen, Yang Gao, Sheng Guo, Jan Hermann, Matthew~R. Hermes, Kevin Koh, Peter Koval, Susi Lehtola, Zhendong Li, Junzi Liu, Narbe Mardirossian, James~D. McClain, Mario Motta, Bastien Mussard, Hung~Q. Pham, Artem Pulkin, Wirawan Purwanto, Paul~J. Robinson, Enrico Ronca, Elvira~R. Sayfutyarova, Maximilian Scheurer, Henry~F. Schurkus, James E.~T. Smith, Chong Sun, Shi-Ning Sun, Shiv Upadhyay, Lucas~K. Wagner, Xiao Wang, Alec White, James~Daniel Whitfield, Mark~J. Williamson, Sebastian Wouters, Jun Yang, Jason~M. Yu, Tianyu Zhu, Timothy~C. Berkelbach, Sandeep Sharma, Alexander~Yu. Sokolov, and Garnet Kin-Lic Chan.
\newblock Recent developments in the pyscf program package.
\newblock {\em J. Chem. Phys.}, 153(2):024109, 2020.
\newblock URL: \url{https://doi.org/10.1063/5.0006074}.

\bibitem{gris+18prl}
Andrea Grisafi, David~M. Wilkins, G{\'a}bor Cs{\'a}nyi, and Michele Ceriotti.
\newblock Symmetry-adapted machine learning for tensorial properties of atomistic systems.
\newblock {\em Phys. Rev. Lett.}, 120(3):036002, 2018.
\newblock URL: \url{https://doi.org/10.1103/PhysRevLett.120.036002}.

\bibitem{Rossi2025}
Mariana Rossi, Kevin Rossi, Alan~M. Lewis, Mathieu Salanne, and Andrea Grisafi.
\newblock Learning the electrostatic response of the electron density through a symmetry-adapted vector field model.
\newblock {\em J. Phys. Chem. Lett.}, 16(9):2326--2332, 2025.
\newblock URL: \url{https://doi.org/10.1021/acs.jpclett.5c00165}.

\bibitem{grisafi2023prm}
Andrea Grisafi, Augustin Bussy, Mathieu Salanne, and Rodolphe Vuilleumier.
\newblock Predicting the charge density response in metal electrodes.
\newblock {\em Phys. Rev. Mater.}, 7(12):125403, 2023.
\newblock URL: \url{https://doi.org/10.1103/PhysRevMaterials.7.125403}.

\bibitem{Grisafi2024}
Andrea Grisafi and Mathieu Salanne.
\newblock Accelerating qm/mm simulations of electrochemical interfaces through machine learning of electronic charge densities.
\newblock {\em J. Chem. Phys.}, 161(2):024109, 2024.
\newblock URL: \url{https://doi.org/10.1063/5.0218379}.

\bibitem{lou2026prx}
Zekun Lou, Alan~M. Lewis, and Mariana Rossi.
\newblock Long-range machine learning of electron density for twisted bilayer moir{\'e} materials.
\newblock {\em PRX Intelligence}, 1(1):013015, 2026.
\newblock URL: \url{https://doi.org/10.1103/4575-9cmx}.

\bibitem{Burke2012persp}
Kieron Burke.
\newblock Perspective on density functional theory.
\newblock {\em J. Chem. Phys.}, 136(15):150901, 2012.
\newblock URL: \url{https://doi.org/10.1063/1.4704546}.

\bibitem{Jones2015rmp}
R.~O. Jones.
\newblock Density functional theory: Its origins, rise to prominence, and future.
\newblock {\em Rev. Mod. Phys.}, 87(3):897--923, 2015.
\newblock URL: \url{https://doi.org/10.1103/RevModPhys.87.897}.

\bibitem{Jorgensen2022}
Peter~Bj{\o}rn J{\o}rgensen and Arghya Bhowmik.
\newblock Equivariant graph neural networks for fast electron density estimation of molecules, liquids, and solids.
\newblock {\em npj Comput. Mater.}, 8(1):183, 2022.
\newblock URL: \url{https://doi.org/10.1038/s41524-022-00863-y}.

\bibitem{Focassio2023}
Bruno Focassio, Michelangelo Domina, Urvesh Patil, Adalberto Fazzio, and Stefano Sanvito.
\newblock Linear jacobi-legendre expansion of the charge density for machine learning-accelerated electronic structure calculations.
\newblock {\em npj Comput. Mater.}, 9(1):87, 2023.
\newblock URL: \url{https://doi.org/10.1038/s41524-023-01053-0}.

\bibitem{Koker2024}
Teddy Koker, Keegan Quigley, Eric Taw, Kevin Tibbetts, and Lin Li.
\newblock Higher-order equivariant neural networks for charge density prediction in materials.
\newblock {\em npj Comput. Mater.}, 10(1):161, 2024.
\newblock URL: \url{https://doi.org/10.1038/s41524-024-01343-1}.

\bibitem{Feng2025}
Chaoqiang Feng, Yaolong Zhang, and Bin Jiang.
\newblock Efficient sampling for machine learning electron density and its response in real space.
\newblock {\em J. Chem. Theory Comput.}, 21(2):691--702, 2025.
\newblock URL: \url{https://doi.org/10.1021/acs.jctc.4c01355}.

\bibitem{Cangi_MALA_2025}
Attila Cangi, Lenz Fiedler, Bartosz Brzoza, Karan Shah, Timothy~J. Callow, Daniel Kotik, Steve Schmerler, Matthew~C. Barry, James~M. Goff, Andrew Rohskopf, Dayton~J. Vogel, Normand Modine, Aidan~P. Thompson, and Sivasankaran Rajamanickam.
\newblock Materials learning algorithms (mala): Scalable machine learning for electronic structure calculations in large-scale atomistic simulations.
\newblock {\em Comput. Phys. Commun.}, 314:109654, 2025.
\newblock URL: \url{https://doi.org/10.1016/j.cpc.2025.109654}.

\bibitem{bigi_metatensor_2026}
Filippo Bigi, Joseph~W. Abbott, Philip Loche, Arslan Mazitov, Davide Tisi, Marcel~F. Langer, Alexander Goscinski, Paolo Pegolo, Sanggyu Chong, Rohit Goswami, Pol Febrer, Sofiia Chorna, Matthias Kellner, Michele Ceriotti, and Guillaume Fraux.
\newblock Metatensor and metatomic: Foundational libraries for interoperable atomistic machine learning.
\newblock {\em J. Chem. Phys.}, 164(6):064113, 2026.
\newblock URL: \url{https://doi.org/10.1063/5.0304911}.

\bibitem{grisafi2019jcp}
Andrea Grisafi and Michele Ceriotti.
\newblock Incorporating long-range physics in atomic-scale machine learning.
\newblock {\em J. Chem. Phys.}, 151(20):204105, 2019.
\newblock URL: \url{https://doi.org/10.1063/1.5128375}.

\bibitem{Rackers2023}
Joshua~A. Rackers, Lucas Tecot, Mario Geiger, and Tess~E. Smidt.
\newblock A recipe for cracking the quantum scaling limit with machine learned electron densities.
\newblock {\em Machine Learning: Sci. Technol.}, 4(1):015027, 2023.
\newblock URL: \url{https://doi.org/10.1088/2632-2153/acb314}.

\bibitem{fu2024recipe}
Xiang Fu, Andrew Rosen, Kyle Bystrom, Rui Wang, Albert Musaelian, Boris Kozinsky, Tess Smidt, and Tommi Jaakkola.
\newblock A recipe for charge density prediction.
\newblock In {\em Advances in Neural Information Processing Systems (NeurIPS)}, 2024.
\newblock URL: \url{https://doi.org/10.52202/079017-0310}.

\bibitem{bogojeski2026}
Mihail Bogojeski, Muhammad~R. Hasyim, Leslie Vogt-Maranto, Klaus-Robert M{\"u}ller, Kieron Burke, and Mark~E. Tuckerman.
\newblock Enhancing molecular dynamics with equivariant machine-learned densities, 2026.
\newblock URL: \url{https://doi.org/10.48550/arXiv.2604.24563}.

\bibitem{numba2015}
Siu~Kwan Lam, Antoine Pitrou, and Stanley Seibert.
\newblock Numba: A llvm-based python jit compiler.
\newblock In {\em Proceedings of the Second Workshop on the LLVM Compiler Infrastructure in HPC}, page~7, 2015.
\newblock URL: \url{https://doi.org/10.1145/2833157.2833162}.

\bibitem{mpi4py2021}
Lisandro Dalcin and Yao-Lung~L. Fang.
\newblock mpi4py: Status update after 12 years of development.
\newblock {\em Comput. Sci. Eng.}, 23(4):47--54, 2021.
\newblock URL: \url{https://doi.org/10.1109/MCSE.2021.3083216}.

\bibitem{fabr+20chimia}
Alberto Fabrizio, Ksenia Briling, Andrea Grisafi, and Clemence Corminboeuf.
\newblock Learning (from) the electron density: Transferability, conformational and chemical diversity.
\newblock {\em Chimia}, 74(4):232--236, 2020.
\newblock URL: \url{https://doi.org/10.2533/chimia.2020.232}.

\bibitem{seifert2026}
Lukas~M. Seifert, Daniel Br{\"u}x, Teodora~M. Piel, and Florian Kleemiss.
\newblock Grid-free quantum crystallographic refinement using unbiased analytic atomic form factors from density fitting.
\newblock {\em Z. Kristallogr. - Cryst. Mater.}, 2026.
\newblock URL: \url{https://doi.org/10.1515/zkri-2026-0013}.

\end{thebibliography}

\end{document}